\documentclass[twocolumn,showpacs,preprintnumbers,amsmath,amssymb]{revtex4}

\usepackage{graphicx}
\usepackage{dcolumn}
\usepackage{bm}

\newcommand{\vf}{v_{\scriptsize{\mbox{f}}}}
\newcommand{\delf}{\delta_{\scriptsize{\mbox{f}}}}
\newcommand{\df}{d_{\scriptsize{\mbox{f}}}}
\newcommand{\dl}{d_{\scriptsize{\mbox{l}}}}
\newcommand{\ds}{d_{\scriptsize{\mbox{s}}}}
\newcommand{\dmin}{d_{\scriptsize{\mbox{min}}}}
\newcommand{\be}{\begin{equation}}
\newcommand{\ee}{\end{equation}}
\newcommand{\lra}[1]{\langle #1 \rangle }
\newcommand{\avg}[1]{\langle #1 \rangle }

\begin{document}
\title{Reaction Spreading on Percolating Clusters}

\author{Federico Bianco}
\affiliation{Dipartimento di Fisica, Universit\`a ``La Sapienza'', Piazzale Aldo Moro 2, I-00185 Roma, Italy}
\affiliation{Institut D'Alembert University Pierre et Marie Curie, 4, place jussieu 75252 Paris Cedex 05}

\author{Sergio Chibbaro}\affiliation{Institut D'Alembert University Pierre et Marie Curie, 4, place jussieu 75252 Paris Cedex 05}\affiliation{CNRS UMR 7190, 4, place jussieu 75252 Paris Cedex 05}

\author{Davide Vergni}\affiliation{Istituto Applicazioni del Calcolo, CNR, Viale Manzoni 30, 00185, Rome, Italy}

\author{Angelo Vulpiani}\affiliation{Dipartimento di Fisica, Universit\`a ``La Sapienza'' and ISC-CNR, Piazzale Aldo Moro 2, I-00185 Roma, Italy}

\begin{abstract} 
Reaction diffusion processes in two-dimensional percolating structures are
investigated. Two different problems are addressed: reaction spreading
on a percolating cluster and front propagation through a percolating channel.
For reaction spreading, numerical data and analytical estimates
show a power law behaviour of the reaction product as $M(t)\sim t^{\dl}$,
where $\dl$ is the connectivity dimension. In a percolating channel,
a statistically stationary
travelling wave develops. The speed and the width of the travelling
wave are numerically computed. While the front speed is a low-fluctuating
quantity and its behaviour can be understood using simple theoretical
argument, the front width is a high-fluctuating quantity showing
a power-law behaviour as a function of the size of the channel.
\end{abstract}

\maketitle
\section{Introduction}
Reaction diffusion processes have been extensively studied in the
past years as systems able to shed some light on various problems of
different disciplines~\cite{Murray, Peters}.  Recently, the importance 
of the non-homogeneity of the medium over which the reaction 
and diffusion take place has been highlighted~\cite{Porto1997}, 
since the qualitative
and quantitative features of the spreading of the reaction process can
depend on the presence of system irregularities. Many studies in the
last years concern reaction/diffusion process in heterogeneous media
accomplishing different problems spacing from epidemic evolution in
heterogeneous networks~\cite{Vespignani2007}, or the intracellular
calcium dynamics~\cite{Thul2004} to the combustion in porous
media~\cite{Tarta2008}.  In this context, studies on reaction dynamics
on percolating clusters appear very interesting for their physical 
relevance and their applications in many different
scientific and technological fields~\cite{Degennes,cardy,Hav,isich}.
For recent experimental and numerical results for reaction-diffusion
on heterogeneous media, see~\cite{Sol_1,Sol_2,Gor_11,Marin,Das,Atis}

The study of reaction and diffusion dynamics on homogeneous
substrate date back to the Fisher-Kolmogorov-Petrovskii-Piskunov 
(FKPP) model~\cite{FKPP}
\begin{equation}
   \partial_t \theta  =
   D \Delta \theta + \alpha g(\theta)\,,
   \label{eq:rd}
\end{equation}
where the scalar field $\theta$ represents the fractional
concentration of the reaction products, $D$ is the molecular
diffusivity, $g(\theta)$ describes the reaction process and $\alpha$
is the reaction rate, i.e., the inverse of the characteristic time,
$\tau$, of the reaction process. In the original model~\cite{FKPP}
$g(\theta)$ assumes a convex shape $g(\theta)=\theta(1-\theta)$. It is
possible to show that under very general conditions~\cite{FKPP},
i.e. if $g(\theta)$ is a convex function and $g'(0)=1$, a travelling
wave develops with asymptotic speed and width given by
$$v_0=2\sqrt{\alpha D},\,\,\,\,\,\,\,\delta_0=c\sqrt{D/\alpha}$$ where
the constant $c$ depends on the definition adopted for the
computation of the front width.
\\
Afterward, as previously mentioned, reaction-transport
dynamics attracted a considerable interest for their relevance in an
incredible large number of chemical, biological and physical
systems~\cite{Murray,Peters}.  In general, when dealing with a non trivial
environment for the reaction and diffusion process it is possible to
extend Eq.~(\ref{eq:rd}) in order to take into account the properties
of the medium~\cite{acvv01,Mancinelli,bcvv2012}:
\begin{equation}
   \partial_t \theta = 
      \hat{L} \theta + f(\theta)\,\,,
   \label{eq:evol}
\end{equation}
where the linear operator $\hat{L}$ rules the transport process.  An
important class of processes of this type is the
advection-reaction-diffusion processes, where
$\hat{L}=-\mathbf{u} \cdot \mathbf{\nabla}+D \Delta$ (e.g.,
see~\cite{acvv01}).
On the other hand it is possible to extend the $\hat{L}$ operator in
order to include cases of effective diffusion on fractal objects
$\hat{L}=\frac{1}{r^{\df-1}}
\frac{\partial}{\partial r}\left(k(r)r^{\df-1}
\frac{\partial}{\partial r}\right)$~\cite{Procaccia1985} 
suitable to study reaction dynamics on fractals~\cite{Mendez2010}.
Moreover in a recent paper~\cite{bcvv2012}, the reaction spreading on
graphs has been considered; in such a case, the operator $\hat{L}$ is
nothing but the Laplacian operator for graphs~\cite{bollobas1998,bc05}. 
In the present paper, in the spirit of the cited works, we study
reaction and diffusion dynamics on percolation clusters, considering
the spreading properties of such a process. \\
In Sect.~2 we present the model and some numerical details. Sect.~3
is devoted to the study of reaction spreading in a large percolating
cluster, while front propagation in a percolating channel is
discussed in Sect.~4. In Sect.~5 the reader can find some conclusions.

\section{Model}
A natural model to study reaction and diffusion on a two dimensional 
non homogeneous medium can be constructed starting from a generalization
of Eq.~(\ref{eq:rd}) in which the transport operator, 
$\hat L=D({\mathbf x}) \Delta $, depends on the spatial variable:
\begin{equation}
   \partial_t \theta({\mathbf x},t) =
   D({\mathbf x}) \Delta \theta({\mathbf x},t) + f(\theta({\mathbf x},t))\,.
   \label{eq:nhomrd}
\end{equation}
The shape and the spatial distribution of $D({\mathbf x})$ permits to
take into account the properties of the medium and therefore to
consider different physical and biological topics~\cite{SK_97,OL_01}.  
In this way it is
possible to study the reaction dynamics at the ``microscopic'' level
without assuming any effective equation able to incorporate mainly
qualitative features of the heterogeous 
medium~\cite{Procaccia1985, Mendez2010, Alonso2009}. 
\\
Since we are mainly interested on the scaling properties 
of the asymptotic behaviour of the system,
without weakening the results we consider the case in which
the variable $D(x)$ can assume only two values, i.e., $D(x)=0$ in
forbidden spatial regions and $D(x)=D_0$ in permitted ones.
The second step is to consider a spatial discretization of
Eq.~(\ref{eq:nhomrd}). The spatial region under examination has been
discretized using a 2d Euclidean lattice,
$\mathcal{L}$, where $\Delta x$ is the lattice constant.  Points
${\mathbf x}$ are replaced by sites of the lattice $s=(i,j)$.

The percolating clusters have been obtained as follows.  Each site may
be permitted (with probability $p$) or prohibited (with probability
$1-p$). If $p>p_c$, where $p_c \simeq 0.592746$ is the site
percolation threshold for square lattices, there is a good chance that
the reaction, starting from any of the permitted sites can invade the
system (percolation).  We call $\mathcal{P}$ the set of the permitted
sites.  In each permitted site we have a value of the concentration
field, $\theta_s(t)=\theta_{(i,j)}(t)$. Eq~(\ref{eq:nhomrd}) can be
discretized as follows
\begin{equation}
   \frac{\mathrm d}{\mathrm dt} \theta_s = \sum_{s'}C_{s,s'}\theta_{s'}
   + f(\theta_{s})\,,
   \label{eq:disnhomrd}
\end{equation}
where $\sum_{s'}C_{s,s'}\theta_{s'}$ is the discretization of
the general transport operator $\hat{L}=D(x) \Delta \theta({\mathbf
x},t)$.  Since we are working on a discrete structure the value
of the lattice spacing $\Delta x$ is not particularly important
(it can be ``absorbed'' in $D_0$ for $D_0$ and $\delta_0$ large enough),
therefore we assume $\Delta x=1$.

In order to specify the quantity $C_{s,s'}$ we introduce the variable
$A_s$ that characterizes the permitted region of the lattice:
\begin{equation}
A_s=\left\{
\begin{array}{cl}
1 & {\rm if } \ s \in \mathcal{P} \cr
0 & {\rm if } \ s \not\in \mathcal{P} \cr
\end{array}
\right .
\label{defLattice}
\end{equation}
Given a site $s$ we can define $k_s=\sum_{|s'-s|=1}A_{s'}$ as the
number of permitted nearest neighbors of $s$.
Using these quantities and imposing the mass conservation of the
diffusion operator, we can express $C_{s,s'}$ as
\begin{equation}
C_{s,s'}=D_0\left\{
\begin{array}{ll}
0 & {\rm if } \,\,\,\, |s-s'|>1 \cr
A_{s'} & {\rm if } \,\,\,\, |s-s'|=1 \cr
k_s & {\rm if } \,\,\,\,s=s'
\end{array}
\right .
\label{defCS}
\end{equation}
So that, the discretized transport term
$\hat{L}\theta_s=\sum_{s'}C_{s,s'}\theta_{s'}$ becomes the discrete
Laplacian of the lattice~\cite{bollobas1998}:
\begin{equation}
\hat{L}\theta_s(t)=
   D_0\left(\sum_{|s'-s|=1}\left(A_{s'}\theta_{s'}(t)\right)-k_s\theta_s(t)
      \right)
\label{defL}
\end{equation}

Finally, the complete model of reaction diffusion on percolating clusters reads
\begin{equation}
\frac{\mathrm d}{\mathrm dt} \theta_s = D_0 \left (
       \sum_{|s'-s|=1}\left(A_{s'}\theta_{s'}(t)\right)-k_s\theta_s(t)
       \right ) + \alpha g(\theta_s)\,\,,
   \label{eq:model}
\end{equation}
where, following classical works~\cite{FKPP} we choose 
$f(\theta)=\alpha g(\theta)$ where $g(\theta)= \theta_s(1-\theta_s)$.
From the numerical viewpoint, given the spatial discretization, 
the temporal derivative is computed via a 4-th order Runge-Kutta algorithm.

\begin{figure}[!h]
\begin{center}
\includegraphics[scale=0.5]{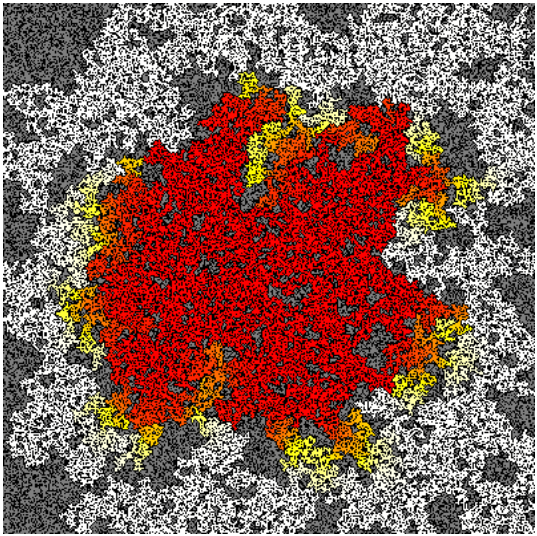}
\caption{Reaction spreading on a square lattice (color online). 
The red area contains reaction products, the yellow area
is the one where the reaction takes place, and the white area
contains fresh material. The black dots indicate
prohibited sites, whereas the various grey areas indicate
regions of permitted sites that do not belong to the percolating cluster 
(we call them islands).}
\vspace{-0.5truecm}
\label{fig:1}
\end{center}
\end{figure}

In the following we study two different problems.  The first concerns
reaction spreading on a large 2d percolating cluster without a
specific geometry (see Figure~\ref{fig:1}), and starting from an initial
condition $\theta_s(0)=0$ except a single site, $\tilde s$, in which
$\theta_{\tilde s}(0)=1$.  In the second problem we study the front
propagation features (speed and width of the travelling wave) in a 2d
channel with dimensions $L_x \times L_y$ with $L_x \geq L_y$ (see
Figure~\ref{fig:3}). In the numerical computations, the lattice
is dynamically modified in order to follow the reacting front, i.e.,
the domain considered in the computation moves rigidly downstream 
when in the upstream part the reaction is extinguished. 
In all the simulations, without lack of generality, we fix $D_0=1$.

It is worth saying that, for $p<p_c$, the propagation 
is practically forbidden if the system is very large.
For finite systems one has yet a possible propagation
if $p$ is not much lower than $p_c$, as can be seen
in the following (in particular in Fig~\ref{fig:4}).

\section{Reaction spreading}
An important quantity that characterizes the
spreading of the reaction is the total mass 
of the reaction product, i.e.,
$M(t)=\sum_{s\in \mathcal{P}}\theta_s(t)\,.$

Since this quantity depends on the number of permitted sites,
we introduce the percentage of $m(t)$ over the lattice, i.e.,
\begin{equation}
m(t)=\frac{M(t)}{N}=\frac{\displaystyle{\sum_{s\in \mathcal{P}}\theta_s(t)}}
                         {\displaystyle{\sum_{s\in \mathcal{L}}A_s}}
\label{eq:obs}
\end{equation}
where $N=\sum_{s\in \mathcal{L}}A_s=\sum_{s\in \mathcal{P}}1$ 
is the number of permitted sites.

Let us briefly remind some relevant quantities in the statistical
analysis of generic graphs: the fractal dimension, $\df$, the
connectivity dimension, $\dl$, (also called chemical dimension) and
the spectral dimension, $\ds$.  The fractal dimension 
$\df$~\cite{ccv} describes the scaling of the number of 
permitted sites in a sphere of radius $r$ in the lattice, as ${\mathcal
N}(r)\sim r^{\df}$.  The connectivity dimension,
instead, measures the average number of sites connected to a given site
in at most $l$ step, as ${\mathcal N}(l) \sim l^{\dl}$.  The spectral
dimension is related to diffusion processes on graphs and can be
defined in terms of the return probability $P_{ii}$ at site $i$ for a
random walker by $P_{ii}(t)\sim t^{-\ds/2}$, or equivalently in terms
of the density of eigenvalues of the Laplacian operator~\cite{bc05}.
The connectivity and fractal dimension can be obviously
different and they are related via the mapping between the two
distances $r$ and $l$~\cite{Havlin1984}. In particular for site
percolation in square lattices, the case of the present study,
at percolation threshold $p\sim p_c$ one has $\df \simeq 1.896$ 
but $\dl \simeq 1.67$ (and, for completeness, $\ds \simeq 1.36$).

\begin{figure}[!h]
\begin{center}
\includegraphics[scale=0.45]{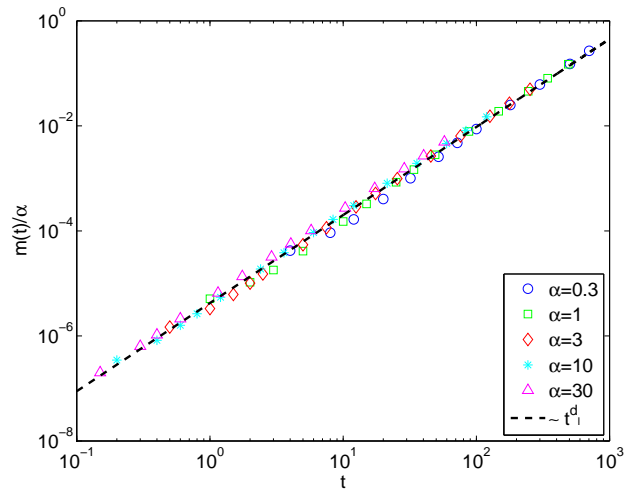}
\caption{Time evolution at $p=0.595\approx p_c$ of the percentage of 
the quantity of 
products rescaled by $\alpha$, $m(t)/\alpha$ vs $t$, 
together with the prediction $m(t)/\alpha \simeq t^{\dl}$
with $\dl\simeq 1.67$ (dashed line). In order to get smooth quantities, 
$m(t)$ is averaged over many realizations ($\approx 5000$) of lattices
of size $L_x=1000$ and $L_y=1000$ which are portions of larger lattices.}
\vspace{-0.5truecm}
\label{fig:2}
\end{center}
\end{figure}

Which is the right quantity that characterizes the reaction spreading?
Numerical computations in agreement with analytical
arguments~\cite{bcvv2012} suggest that the chemical dimension is the
right quantity. Starting from a single site with $\theta_i(0)=1$,
after $t$ step the number of site reached by the field is ${\mathcal
N}(t)\sim t^{\dl}$~\cite{Grass}. Therefore, in the limit of very fast 
reaction, when each site reached by the field is immediately burnt 
(i.e, $\theta_s \simeq 1$), we can expect:
\begin{equation}
m(t)\sim t^{\dl}.
\label{eq:dymobs}
\end{equation}
Fig.~\ref{fig:2} clearly shows the scaling of Eq.~(\ref{eq:dymobs}).
Moreover this Figure reveals that the scaling~(\ref{eq:dymobs})
is valid not only in the fast reaction regime, and that the reaction rate 
is relevant only for the prefactor: $m(t)\simeq\alpha t^{\dl}$.

\section{Front propagation}
The problem of the front propagation in reactive systems (classical
reaction and diffusion processes, advection reaction and diffusion 
processes, reaction and diffusion in the presence of anomalous diffusion, 
etc.) has been extensively studied~\cite{Murray, Peters, Mancinelli}.  
In some cases, under certain conditions, it is possible to show that
the propagation is standard, i.e., there exists an asymptotic value
for the speed $v$ and the width $\delta$ of the propagating front.
On the other hand, it is pretty impossible
(except very special cases) to determine analytically the
values of $v$ and $\delta$. Therefore the numerical study of the speed
and the thickness of the moving front is mandatory to obtain
information about the spreading dynamics.

In the case of reaction processes on percolating clusters, if one
considers an arbitrarily large (in any direction) lattice, the propagation
generally is not standard since the total quantity of reaction products
grows as a power law with a non integer exponent, $m(t)\simeq\alpha
t^{\dl}$.  If the percolating cluster is embedded in a channel 
with a propagation direction, $L_x$, and a transversal direction, $L_y$, 
with $L_x \geq L_y$, a travelling wave takes place with a constant 
(on average) speed after a transient needed to the reaction product to
invade the transversal direction of the channel. Therefore, we consider 
the model~(\ref{eq:model}) with an initially empty 2d lattice where 
$L_x \geq L_y$ and $\theta_{(i,j)}(0)=0$. In order to reduce the transient, 
as boundary conditions we use $\theta_{(i=0,j)}(t)=1$ and
$\theta_{(i=L_x,j)}(t)=0$ for the left and right edge, respectively.
In the transversal direction we have zero-flux (Neumann) boundary
conditions, that are automatically guaranteed by the diffusion operator
(\ref{defL}). Using the above boundary conditions, we expect the
development of a front propagating with a fixed (on average)
speed from the left to the right side of the lattice.

\begin{figure}[!h]
\begin{center}
\hspace{0.6truecm}\includegraphics[scale=0.1713]{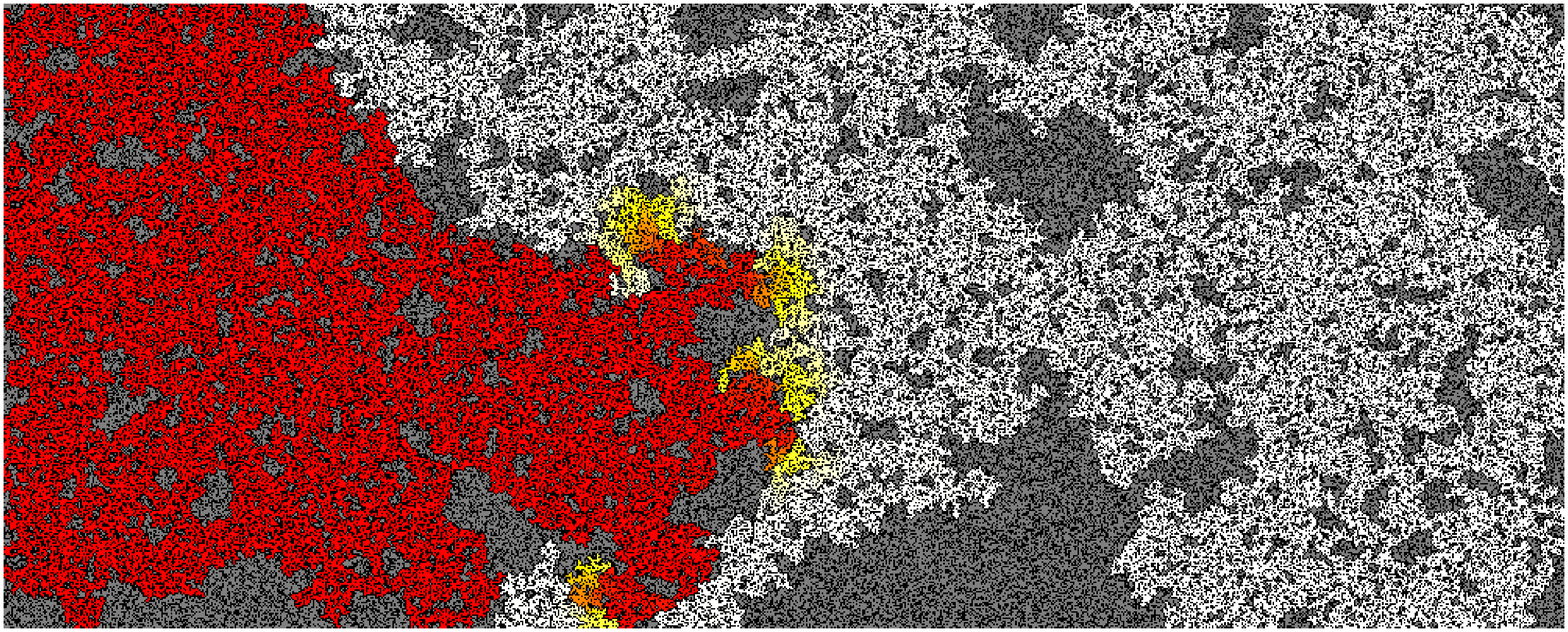}\\
\includegraphics[scale=0.6]{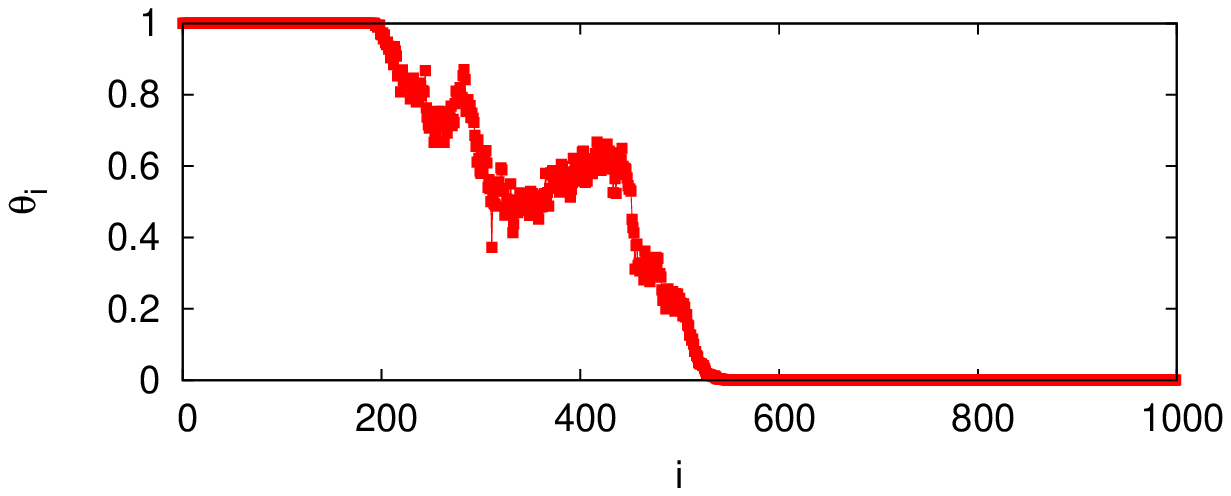}
\caption{Snapshot of the reaction dynamics in a percolating cluster
on a channel (color online, see caption of Figure 1). The graph below
shows the profile of the average of the front (see Eq.~(\ref{eq:avfield}))
related to the snapshot.}
\vspace{-0.5truecm}
\label{fig:3}
\end{center}
\end{figure}

In Fig.~\ref{fig:3} it is shown an example of front propagation in
a percolating cluster.  The dynamic evolves through the horizontal
direction with a fluctuating front depending on the position of the
permitted sites. Because of such fluctuations it is convenient to 
introduce the averaged field along the horizontal direction as the
mean of the field $\theta_s(t)$ along the $i$-direction
\begin{equation}
\theta_i(t) = \frac{\displaystyle{\sum_{j=1}^{L_y}A_{(i,j)}\theta_{(i,j)}(t)}}
                   {\displaystyle{\sum_{j=1}^{L_y}A_{(i,j)}}}.
\label{eq:avfield}
\end{equation}

Strictly speaking, given a percolating cluster, the moving front is
not a travelling wave in the classical sense, since there does not exist a
function $f(i)$ such as $\theta_i(t)=f(i-vt)$. This is due both for the 
random nature of the permitted sites on the lattice (i.e., the average
stabilizes only at very large $L_y$) and for the discrete nature
of the lattice. But it is still possible to define averaged quantities
such as the propagation speed or the front width as follows.

\subsection{Front speed}
In the case of travelling waves, we expect that the total mass
of the reaction products increases, on average, linearly with time
\begin{equation}
M(t)\simeq  N_p v t
\label{eq:evM}
\end{equation}
where $N_p$ is the averaged number of site accessible by the reaction
process in the vertical direction.  The computation of $N_p$ is a
quite delicate point thus the maximum amount of accessible sites in
a single column of the lattice is not $L_y$ (since there are permitted
and prohibited site in the lattice) neither $pL_y$ (since not the
whole set of permitted sites belong to the percolating cluster). 
$N_p$ can be estimated as follows.  At the percolating threshold, $p\sim
p_c$, the total number of points belonging to the percolating cluster in a
square of size $L_y$ is proportional to $L_y^{\df}$. Since there are
$L_y$ rows in the square one has $N_p \sim
\frac{L_y^{\df}}{L_y}=L_y^{\df-1}$.
Instead, if $p$ is large enough to
have one single big percolating cluster, without the presence of
closed islands of permitted sites not connected to the principal
percolating cluster, one has $N_p \simeq p L_y$. In the intermediate 
cases it is possible to compute $N_p$ numerically. Therefore, we can
define the average front speed as
\begin{equation}
v_1=\lim_{t\to\infty}\frac{M(t)}{N_p t}\,.
\label{eq:defv1}
\end{equation}
Another way to define $v$, that is much more sensitive to statistical 
fluctuations of the cluster structure, can be obtained starting from the 
dynamics of the model.
Since the diffusion operator~(\ref{defL}) is a mass-preserving term,
the derivative of the total mass can be computed using Eq.~(\ref{eq:model}) 
\begin{equation}
v_2(t)=\frac{1}{N_p}\frac{{\mathrm d}}{{\mathrm d}t} M(t) = 
\frac{\alpha}{N_p} \sum_{s\in \mathcal{P}}(\theta_s(t)(1-\theta_s(t)))\,.
\label{eq:defv2}
\end{equation}
Of course $v_2(t)$ is a function of time and its fluctations reflect the
random nature of the percolating cluster. On the other hand, we expect, 
as confirmed from numerical simulation (not shown), that 
$\avg{v_2(t)}=v_1=\vf$. 

\begin{figure}[!h]
\begin{center}
\includegraphics[scale=0.5]{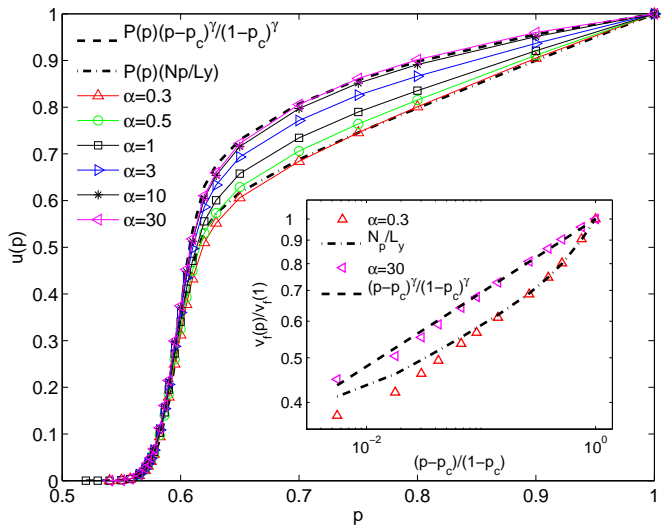}
\caption{Average front speed $u(p)$ as a function of $p$ 
for various $\alpha$ together with asymptotic 
behaviour~(\ref{eq:asym1}) and~(\ref{eq:asym2})
(with $\gamma\simeq 0.16$). The channel length 
is taken to be $L_x=100$ and $L_y=100$.
In the inset it is shown in log-log scale the
behaviour of $\vf(p)/\vf(1)$ together with
the theoretical prediction: 
$\vf(p)/\vf(1)\sim N_p/L_y$ for slow
reaction rate and $\vf(p)/\vf(1)\sim ((p-p_c)/(1-p_c))^\gamma$
for fast reaction rate.}
\vspace{-0.5truecm}
\label{fig:4}
\end{center}
\end{figure}

It is interesting to study the behaviour of $\vf$ as a function of 
$p$, the probability of having a permitted site. 
In fact, using different values of $p$ it is possible to model 
different degree of non-homogeneity and we expect different
evolution of the reaction process. For $p=1$, since the 
lattice is homogeneous, we expect to obtain the FKPP value 
$v_0=2\sqrt{\alpha D_0}$. This result is true for small $\alpha$,
when $\delta_0\sim\sqrt{D_0/\alpha}$ is larger than the lattice size 
(simulations not shown for the sake of brevity).
On the contrary for large $\alpha$, because
of the discrete nature of the lattice, the width of the FKPP front 
can be of the same order, or even smaller, of the lattice step.
In this case there is a significant difference between the measured 
front speed and the FKPP value also for $p=1$. 
Although this discrepancy does not invalidate our analysis, 
we choose to study only rescaled velocity $\vf(p)/\vf(1)$.
 
In the case of $p < 1$, especially for $p \sim p_c$, it is important
to introduce the probability of having a percolating lattice, $P(p)$. 
We write $u(p)=P(p)\vf(p)/\vf(1)$ as the
average velocity in a percolating cluster when the site probability is
$p$.  For value of $p$ larger than $p_c$ it is possible to give simple
but valid arguments to explain the behaviour of $u$.  First of all,
for small $\alpha$ values we expect a large front that regularizes the
propagation. This is a kind of homogenization regime~\cite{acvv01}. 
Practically, we can imagine the front proceeding almost as
in a homogeneous medium excluding the region in which the propagation
is prohibited. Therefore we can write 
\begin{equation}
u(p)=P(p)\frac{\vf(p)}{\vf(1)}\sim
P(p)\frac{N_p}{L_y}\,.
\label{eq:asym1}
\end{equation}
Such a relation, when $p$ is large, simplifies to
$\vf(p)\sim p \vf(1)$.  

\begin{figure}[!h]
\begin{center}
\includegraphics[scale=0.5]{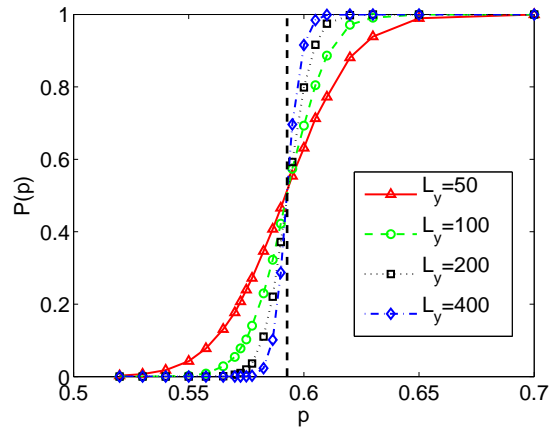}
\caption{The probability $P(p)$ to percolate along a channel 
of size $L_x=L_y$ is shown for different $L_y$.}
\vspace{-0.5truecm}
\label{fig:5}
\end{center}
\end{figure}

In the other limit, for large $\alpha$, we can use the following 
argument~\cite{Havlin2000}.  
We know (from Eq.~(\ref{eq:dymobs})) that $m(t)\sim t^{\dl}$.  
On the other hand $m(t)\sim r(t)^{\df}$. Therefore $r(t)\sim t^{\dl/\df}$, 
and $v=\frac{dr}{dt} \sim t^{\dl/\df-1}\sim r^{1-\dmin}$, where 
$\dmin=\frac{\df}{\dl}$. Furthermore, if the linear size of the region 
is $r<\xi$, where $\xi$ is the correlation length~\cite{Hav},
the cluster is self-similar and then $v \sim \xi^{1-\dmin}$. 
Moreover, analysis of the percolation phase transition gives 
$\xi\sim |p-p_c|^{-\nu}$, with $\nu=4/3$ for $d=2$~\cite{den}, 
which gives the final scaling $v\sim(p-p_c)^\gamma$,
where $\gamma=-\nu(1-\dmin)$.
For the average velocity, the scaling is:
\begin{equation}
u(p)=P(p)\frac{\vf(p)}{\vf(1)}\sim P(p)
\left(\frac{p-p_c}{1-p_c}\right)^\gamma.
\label{eq:asym2}
\end{equation}
Alternatively, a similar scaling had been derived through 
large deviation theory \cite{Mendez2010}.
Both the above behaviors are
well observed in the numerical simulations, as shown in
Fig.~\ref{fig:4}.
It is worth noting that below the percolation threshold, 
the probability to have a percolating cluster tends to zero 
for a channel long enough, see Fig~\ref{fig:5}. 
Nonetheless, in Fig~\ref{fig:4}, it is possible
to observe a very small velocity $u(p)$ for $p\lesssim p_c$.
This result is basically due to the fact that, for finite size, $P(p)$
is not strictly zero for $p\lesssim p_c$, see Fig~\ref{fig:5}. 

Concerning the probability $P(p)$ of having a percolating lattice
as a function of $p$, in the numerical simulations it is possible
to compute $P(p)$ only for finite values of $L_x$ and $L_y$.
Moreover, in applications the cluster size is finite, and 
$L_y$ can be small. Fig.~\ref{fig:5} shows $P(p)$ for different
values of $L_y$ in the case of $L_x=L_y$. 
Naturally for $L_y \rightarrow \infty$,
$P(p)$ approaches the Heaviside step function $\Theta(p-p_c)$.
Simulations (not reported here) show that for non square lattices $L_x=nL_y$ 
with $n > 1$, while the front speed $\vf$ does not change with $n$,
the probability $P(p)$ is strongly influenced by $n$, if $n$ is large. 
Moreover, in the case of large $n$, also $p_c$ changes, becoming
dependent on both $n$ and $L_y$. 

\subsection{Front width}
For a two dimensional propagating wave in random media, we can define
various different measure of width. One important
measure concerns the averaged width of the front along the propagation
direction. It is the analougous to the front width in the 1d FKPP
traveling wave and measures the region along the x direction in which
the reaction process is active (see Figure~\ref{fig:3}).  
In order to define such a quantity
one can use $\theta_i(t)$, i.e. the average over the 
$i$-direction of the field $\theta_s(t)$, defined in Eq.~(\ref{eq:avfield}).
Yet the averaged quantity $\theta_i(t)$ still suffers 
from large fluctuation so we use a simplified observable able to give 
a good measure of the front width. 
First of all we introduce an auxiliary quantity
\begin{equation}
H_{(i,j)}(t)=\left\{
\begin{array}{ll}
1 & {\rm if } \,\,\,\, 0.01 \leq \theta_{(i,j)}(t) \leq 0.99 \cr
0 & {\rm elsewhere}
\end{array}
\right .\,.
\label{discretefield}
\end{equation}

\begin{figure}[!h]
\begin{center}
\includegraphics[scale=0.4]{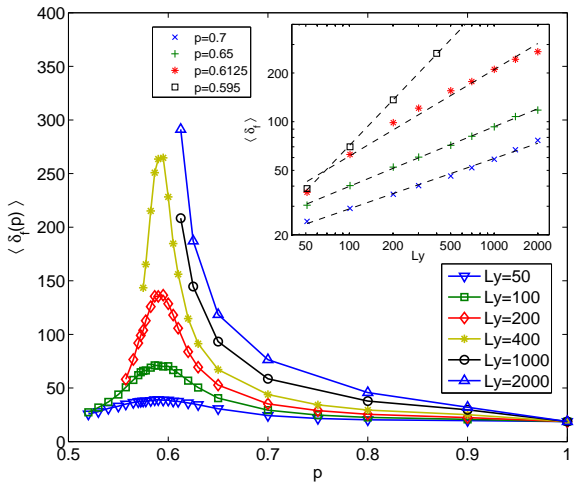}
\caption{Averaged front width, $\lra{\delf}$, as a function of $p$ 
at varying $L_y$, where the horizontal size of the lattice  
is $L_x=5000L_y$. For all curves, $\alpha=1$. In the inset
it is shown, for fixed values of $p$, the scaling behaviour
of $\lra{\delf}\sim L_y^\beta$, where $\beta=0.94$ for $p=0.595$,
$\beta=0.54$ for $p=0.6125$, $\beta=0.37$ for $p=0.65$, 
$\beta=0.31$ for $p=0.7$.}
\vspace{-0.5truecm}
\label{fig:6}
\end{center}
\end{figure}

Then we define $\delf(t)$, the front width, as the distance between 
the maximum and the minimum value of $i$ such that $H_{(i,j)}(t)=1$.
In this way we define a rectangle of size $L_y\,\times\, \delf(t)$
inside which there is the whole active front.
Also $\delf(t)$ is a strongly fluctuating quantity, 
therefore we study the statistical feature of $\delf(t)$, e.g., 
$\lra{\delf}$, as a function of $p$ and $L_y$. 

\begin{figure}[!h]
\begin{center}
\includegraphics[scale=0.4]{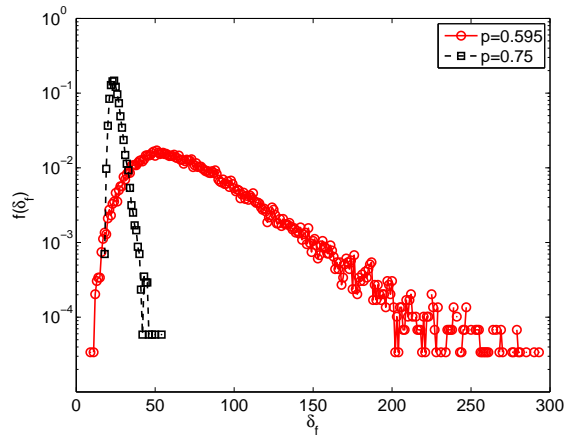}
\caption{Two different probability density functions of ${\delf}$ 
for two values of p are shown: $p=0.595\approx p_c$ and $p=0.75$,
with $L_y=100$, $L_x=5000L_y$ and $\alpha=1$.
It is well evident that for $p\approx p_c$, fluctuations play a dominant 
role and large deviations are present.}
\vspace{-0.5truecm}
\label{fig:7}
\end{center}
\end{figure}

In principle one can expect that given $p$ and $\alpha$, 
for $L_y$ large enough, the front width
reaches a constant value, as for the front speed.  
On the other hand, as shown in Figure~\ref{fig:6}, the convergence 
depends on $p$:
while for large $p$ (near to $p=1$) there is an asymptotic value 
of $\lra{\delf}$,
for values of $p$ going to $p_c$ there is no convergence at all.
Notably, in the limit of very large clusters, the averaged 
front width diverges rapidly around $p\sim p_c$.
As the inset of Fig.~\ref{fig:6} shows, the scaling structure of the
front width as a function of $L_y$ at varying $p$ is highly non trivial
and cannot be associate to a single scaling exponent~\cite{Gross1}.

Rather interesting is the presence of very large fluctuations
of $\lra{\delf}$. Fig.~\ref{fig:7} shows how, for $p$ near $p_c$,
the typical value of the front width, ${\delf}^T$, given
by the maximum of the probability density function,
is of the same order of the fluctuation of $\lra{\delf}$
(measured as $\sqrt{\avg{\delf^2(t)}-\avg{\delf(t)}^2}$).

\begin{figure}[!h]
\begin{center}
\includegraphics[scale=0.5]{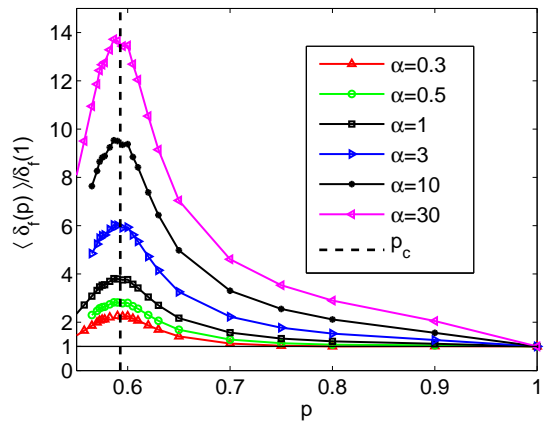}
\caption{Normalised averaged front width for different values of the 
reaction rate with $L_y=100$, $L_x=5000L_y$. 
For small values of $\alpha$ fluctuations are smoothed, 
whereas they are maximal for large reaction rates. }
\vspace{-0.5truecm}
\label{fig:8}
\end{center}
\end{figure}

The above discussion is valid at fixed (and not too small) $\alpha$. 
When $\alpha$
is small, the bare FKPP front width, $\delta_0$, is large,
and a large front width regularizes the reaction dynamics.
If the bare front width is larger than the typical size 
of the prohibited islands (for a given $p$) we can expect 
that the random distribution of the islands does not affect 
too much the front propagation, with a net effect of 
diminishing  the fluctuations and  the dependence
of the front on both $p$ and $L_y$. On the other hand,
for large $\alpha$ the bare front width is comparable
with the lattice discretization. In this case 
fluctuations become very strong as the dependency 
of the front width both from $p$ and $L_y$. Figure~\ref{fig:8}
explicates the above discussion.

\section{Conclusion}
Reaction and diffusion processes in heterogeneous media, because of
their relevance in many real-world applications, play a central role
in several different fields.  In the present paper, starting from the
basic equations, we have investigated the behaviour of a simple
reaction and diffusion process taking place in a heterogeneous medium,
i.e., two-dimensional percolating structures.  We show that for the
reaction spreading on percolating clusters the dynamics is ruled by
the connectivity dimension, $\dl$ (see Eq.~(\ref{eq:dymobs})) and the
reaction rate affects only the prefactor of the scaling. In the case
of percolating clusters through a channel, the reaction and diffusion
process develops a statistically stationary travelling wave.  The
speed and the width of the travelling wave are deeply influenced by
the percolating transition together with finite size effects that
generate peculiar behaviours of both front speed and front
width. Those effects are crucial since, in realistic problems, the
channel over which the reaction takes place has necessarily a finite
transversal length.  Some recent numerical computations and
experiments show the key role played by the flow heterogeneities on
the chemical front dynamics~\cite{Gor_11,Marin,Das,Atis}.

\section{Acknowledgements}
We  thank  R. Burioni for fruitful discussions.


\begin{thebibliography}{99}


\bibitem {Murray} J.D. Murray, {\it Mathematical Biology}, 
(Springer-Verlag, Berlin, 1993).

\bibitem{Peters} N.~Peters, 
{\it Turbulent combustion}
(Cambridge University Press, New York, 2000).

\bibitem{Porto1997}
M. Porto, A. Bunde, S. Havlin, and H.E. Roman, 
Phys. Rev. E {\bf 56}, 1667 (1997).

\bibitem{Vespignani2007} V.~Colizza and A. Vespignani,
Phys. Rev. Lett. {\bf 99}, 148701 (2007).

\bibitem{Thul2004} R.~Thul and M.~Falcke,
Phys. Rev. Lett. \textbf{93}, 188103 (2004).

\bibitem{Tarta2008} A. M. Tartakovsky, D. M. Tartakovsky, 
T. D. Scheibe, and P. Meakin,
SIAM J. Sci. Comput., {\bf 30} 2799 (2008).

\bibitem{Degennes}  P.G. de Gennes, La Recherche, {\bf 7}, 919, (1976); 
P.G. de Gennes, J. Phys. Lett., Paris {\bf 37}, L1, (1976).

\bibitem{cardy} J. L. Cardy and P. Grassberger,
J. Phys. A: Math. Gen. {\bf 18}, L267 (1985).

\bibitem{Hav} A. Bunde and S. Havlin,
{\it Fractals and Disordered Systems}, Springer-Verlag Berlin (1999); 
S. Havlin and D. ben-Avraham,
Advances in Physics, {\bf 36}, 695 (1987).

\bibitem{isich}
M. B. Isichenko, 
Reviews of Modern Physics {\bf 64}, 961 (1992).

\bibitem{Sol_1}
M. S. Paoletti, and T. H.  Solomon  EPL, {\bf 69}, 819 (2005).
\bibitem{Sol_2}
M. E. Schwartz and T. H.Solomon,  Phys. Rev. Lett., {\bf100},
028302 (2008).

\bibitem{Gor_11}
S Goroshin, F-D Tang, and A. J. Higgins, Phys.Rev. E \textbf{84}, 027301(2011). 

\bibitem{Marin} 
\`A. G. Mar\'in, H. Gelderblom, D. Lohse, and J. H. Snoeijer,  Phys. Rev. Lett., \textbf{107}, 085502 (2011).
\bibitem{Das} 
S Das, S Chakraborty, and S.K. Mitra, Phys. Rev. E \textbf{85}, 046311 (2012)
\bibitem{Atis} 
S Atis, S Saha, H Auradou, D Salin, and L Talon, Phys. Rev. Lett., \textbf{110}, 148301 (2013).


\bibitem{FKPP} A.~N.~Kolmogorov, I.~G.~Petrovskii, and N.~S.~Piskunov, 
Moscow Univ. Bull. Math. {\bf 1}, 1 (1937); 
R.~A.~Fischer, Ann. Eugenics {\bf 7}, 355 (1937).

\bibitem{acvv01}
M. Abel, A. Celani, D.Vergni and A. Vulpiani, 
Phys. Rev. E {\bf 64}, 046307 (2001).

\bibitem{Mancinelli}
R. Mancinelli, D. Vergni and A. Vulpiani, 
Physica D {\bf 185}, 175 (2003).

\bibitem{bcvv2012}
R. Burioni, S. Chibbaro, D.Vergni and A. Vulpiani, 
Phys. Rev. E {\bf 86}, 055101 (2012).

\bibitem{Procaccia1985} 
B. O'Shaughnessy, I. Procaccia,
Phys. Rev. Lett. {\bf 54}, 455 (1985);
L. P. Richardson, 
Proc. R. Soc. London A \textbf{110}, 709 (1926).

\bibitem{Mendez2010}
V. Mendez, D. Campos and J. Fort,
Phys. Rev. E {\bf 69}, 016613 (2004);
D. Campos, V. Mendez and J. Fort,
Phys. Rev. E {\bf 69}, 031115 (2004);
V. Mendez, S. Fedotov, and W. Horsthemke, 
{\it Reaction-Transport Systems: Mesoscopic Foundation, Fronts, and Spatial Instabilities}
(Springer-Verlag, Berlin, 2010).

\bibitem{bollobas1998}
B. Bollob\'as, {\it Modern Graph theory}
(Springer-Verlag New York, 1998).

\bibitem{bc05}
R. Burioni and D. Cassi,
J. Phys. A: Math. Gen. {\bf 38}, R45 (2005).

\bibitem{SK_97}	
N. Shigesada and K. Kohkichi. \emph{Biological invasions: theory and practice}, (Oxford University Press, UK, 1997).

\bibitem{OL_01}
A. Okubo and S. A. Levin,  \emph{Diffusion and ecological problems: modern perspectives}. (Springer Verlag, New York, 2001).

\bibitem{Alonso2009} 
S. Alonso, R. Kapral and M. B\"ar,
Phys. Rev. Lett. {\bf 102}, 238302 (2009).

\bibitem{ccv} M Cencini, F Cecconi, and A Vulpiani, 
{\it Chaos},
(World Scientific, Singapore, 2010).

\bibitem{Havlin1984}
S. Havlin and R. Nossal, 
J. Phys. A: Math. Gen. {\bf 17}, L427 (1984);
S. Havlin, D. ben-Avraham,
Adv. Phys. {\bf 36}, 695 (1987);
H. E. Stanley and P. Trunfio,
II Nuovo Cimento D, {\bf 16}, 1039 (1994);
P. Meakin and H. E. Stanley,
J. Phys. A: Math. Gen. {\bf 17}, L173 (1984).

\bibitem{Grass}
P. Grassberger, J. Phys. A: Math. Gen. {\bf 18}, L215 (1985).


\bibitem{den}
M. P. M. den Nijs, J. Phys. A: Math. Gen. {\bf 12}, 1857 (1979); 
B. Nienhuis, Phys. Rev. Lett. {\bf 49}, 1062 (1982).

\bibitem{Havlin2000}
D. ben-Avraham, S. Havlin,
{\it Diffusion and reactions in fractals and disordered systems} 
(Cambridge University Press, New York, 2000).

\bibitem{Gross1} 
T. Grossman and A. Aharony, J. Phys. A: Math. Gen. {\bf 19}, L745 (1986); 
R. F. J. Voss, J. Phys. A: Math. Gen. {\bf 17}, L373 (1984); 
P. Grassberger, J. Phys. A: Math. Gen. {\bf 19}, L2675 (1986); 
H. Saleur and B. Duplantier, Phys. Rev. Lett. {\bf 58}, 2325 (1987).



%
%

\end{thebibliography}
\end{document}